# Low-Energy Astrophysics

## Stimulating the Reduction of Energy Consumption in the Next Decade

P.J. Marshall,[1,2] N. Bennert,[1] E.S. Rykoff,[1,2] K.J. Shen,[1] J.D.R. Steinfadt,[1] J. Fregeau,[1,3] R-R. Chary,[4,5] K. Sheth,[4,5] B. Weiner,[6] K.B. Henisey,[1] E.L. Quetin,[1] R. Antonucci,[1] D. Kaplan,[3] P. Jonsson,[7] M.W. Auger,[1] C. Cardamone,[8] T. Tao,[1] D.E. Holz,[9] M. Bradac,[1] T.S. Metcalfe,[10] S. McHugh,[1] M. Elvis,[11] B.J. Brewer,[12] T. Urrutia,[4,5] F. Guo,[1] W. Hovest,[13] R. Nakajima,[14] B.-Q. For,[15] D. Erb,[1] D. Paneque[16]

[1]UC Santa Barbara, [2]LCOGT, [3]KITP, [4]Caltech, [5]IPAC, [6]Univ. Arizona, [7]UC Santa Cruz, [8]Yale Univ., [9]Los Alamos, [10]NCAR, [11]SAO/Harvard, [12]Univ. Sydney, [13]MPA (Germany), [14]UC Berkeley, [15]Univ. Texas Austin, [16]KIPAC

*An independent submission to Astro2010, the Astronomy and Astrophysics Decadal Survey*

*March 15th 2009*

*In this paper we address the consumption of energy by astronomers while performing their professional duties. Although we find that astronomy uses a negligible fraction of the US energy budget, the rate at which energy is consumed by an average astronomer is similar to that of a typical high-flying businessperson. We review some of the ways in which astronomers are already acting to reduce their energy consumption. In the coming decades, all citizens will have to reduce their energy consumption to conserve fossil fuel reserves and to help avert a potentially catastrophic change in the Earth's climate. The challenges are the same for astronomers as they are for everyone: decreasing the distances we travel and investing in energy-efficient infrastructure. The high profile of astronomy in the media, and the great public interest in our field, can play a role in promoting energy-awareness to the wider population. Our specific recommendations are therefore to 1) reduce travel when possible, through efficient meeting organization, and by investing in high-bandwidth video conference facilities and virtual-world software, 2) create energy-efficient observatories, computing centers and workplaces, powered by sustainable energy resources, and 3) actively publicize these pursuits.*

---

Primary author contact details: Phil Marshall, email: pjm@physics.ucsb.edu, tel: +1 805 893 3189

# 1. Introduction

Like all workers, we astronomers consume energy while at work. A report on the state of any profession would be incomplete without some consideration of its energy consumption: the way the world uses energy is precipitating a global environmental and economic crisis, the threatening ramifications of which are now coming to light. These include:

- **Potentially catastrophic climate change**
- **The depletion of fossil fuel reserves (and the resulting economic and political upheaval)**

The UN Intergovernmental Panel on Climate Change (2007) has recently warned of "large-scale climate events [that] have the potential to cause very large impacts,"[1] and current research projects a 10% chance of devastating warming of more than 7$^o$C over this century if no action is taken.[2] The 2007/2008 UN Human Development Report sees climate change as "the greatest challenge facing humanity at the start of the 21st Century," and notes that "failure to meet that challenge raises the spectre of unprecedented reversals in human development." [3] In 2004, the AAS called for "peer-reviewed climate research to inform climate-related policy decisions, and, as well, to provide a basis for mitigating the harmful effects of global change."[4] Many scientists, including the authors of this paper, have come to the conclusion that the mitigation of the harmful effects of climate change needs to start as soon as possible. It is clear that this mitigation must necessarily involve rapid and severe reductions in our production of greenhouse gases, the most crucial of which being the carbon dioxide produced by burning fossil fuels.[1]

Finding alternatives to fossil fuels has other benefits. Oil, for example, is a finite resource highly valued for more than just its combustion: it is the raw material for plastics, pharmaceuticals, coatings, reagents, and many other products upon which we depend daily. Conserving oil addresses this issue, and both of those given above, but meeting this challenge will require a rethinking in the way we live and work, and an active pursuit of wiser alternatives. Let us briefly put this challenge into perspective by assessing the total energy consumption of Americans, and what it would take to supply them with sustainable energy from the brightest astronomical source, the Sun.

```
For the United States, per capita energy consumption has been estimated as 250 kWh/day/person,[5][6]
including household electricity use, manufacturing, and transportation. Suppose we could supply all
of this energy using electricity from solar power generators, which we might expect to provide
30 W/m^2 on average.[5]   How much sunny desert land would we need?
For 300 million Americans, the answer is:
   3e8 x 250 kWh/day/person / 24h/day / 0.03 kW/m^2 = 1.0e11 m^2
which is 40,000 square miles, or about a third of the area of the state of Arizona.
```

This calculation is not a comment on the feasibility of solar power generation: it is simply an echo of the conclusion of MacKay (2008)[5] and others that sustainable energy generation for an entire nation requires national-scale infrastructure. **The inevitable change to sustainable energy will be expensive, but the less we use, the less we will have to spend.**

This is a paper about the rate of energy consumption in astronomy, and the need for astronomers to be seen to reduce it. We will estimate how much energy we, the US astronomical community, use, and show that our contribution to the total US energy consumption rate is negligible. We will also estimate how much energy we use in our work *per astronomer,* and show approximately how astronomers compare with businesspeople in this regard. All of us in the US must act to reduce our energy consumption; the big, important things that astronomers in the US can and should do are much the same as the things that everyone else can and should do - the key difference is that we astronomers can do them in the public eye. The fact that astronomy is a high-profile activity, receiving more media attention than most (if not all) branches of science, means that the US astronomy community can and should play a leading role in the reduction of energy consumption. As we shall see, astronomers have already begun this work.

## 2. The Energy Consumption of the Average Astronomer

In this section we make a very rough estimate of the energy budget of a professional astronomer. Note the emphasis on calculating energy consumption per day per person; this is the relevant quantity for any exercise in energy accounting. We use the size of the membership of the AAS (7000 members[7]) as a rough estimate of the total number of astronomers in the US.

### 2.1 Non-astronomical Energy Consumption

We can approximate the non-astronomical energy consumption by the energy consumption of the average American citizen. A rough but plausible energy breakdown for a typical American, extrapolating and adjusting from MacKay (2008)[5] is as follows:

```
~90 kWh/day for driving, estimated for a typical 20mpg car (using about 2.7kWh/mile),
            and driving 12000 person-miles/year.[8][9].
~50 kWh/day for heating and cooling your home and workplace.
~20 kWh/day for flying, budgeting 4000 miles/year.
            Traveling by airplane uses about 1.8kWh/mile, per person.
 ~4 kWh/day for lighting,
 ~5 kWh/day for gadgets such as cell-phones, TVs etc.
~15 kWh/day for the growing and fertilization of food.
~50 kWh/day for mining, producing, using, and disposing of "stuff", including packaging, computers,
            houses, roads, etc.
~12 kWh/day for transporting that stuff in trucks, trains, and on the high seas.
 ~2 kWh/day for national defense.

This sums to ~250 kWh/day/person total, which is approximately the same as the estimated per capita
consumption of the US.[3]
```

This gives us a useful reference point, and shows the areas where the average citizen might look to reduce their energy consumption. In addition to this baseline, how much energy do astronomers use just by doing astronomy? To answer this question, we explore a very simplistic toy model of energy consumption in astronomy.

### 2.2 Small-scale, Computationally-intensive Office Work

Most of what we do at work is very similar to that done in other office jobs. The main difference is that we use more computing power than the average office worker.

```
Averaging roughly over desktops (100-150W), laptops (<100W), and 8-core machines (200-400W),
including intermittent monitor use (150W), and accounting for both idle and active periods,
we estimate that local computing requires roughly 150W per astronomer, or 4kWh/day/astronomer.
```

**2.3 Large Facilities**

Let us now consider the energy consumption of large astronomical facilities. In our toy model we consider two classes of facility: observatories, and large-scale computing centers.

**Observatories**: Suppose that the US community were to operate one 30-m class and ten 8-m class optical telescopes, and two VLA-scale radio arrays. We restrict ourselves to these two types of ground-based telescope, due to the large uncertainties in estimating the energy consumption of space and airborne observatories (the latter of which is likely to be substantial[10]). This cartoon picture may not turn out to represent accurately the observing facilities in use during the next decade, but again it serves to give us a simple reference point.

The principal uses of energy at ground-based optical observatories are generally pumps, compressors, and chillers used in the telescope bearing and mirror support systems, and in the thermal control of mirror, telescope, and dome to preserve stable image quality. Computing and other electrical uses also contribute but are relatively smaller. An 8-m class telescope requires of order 150 kW of power averaged over a year and day/night. The number of toy model observatories (ten) was chosen to represent the many smaller telescopes operated in parallel with the existing 8-10m class facilities. A 20 to 30-m class telescope could require 3 to 5 times more.

```
The toy optical observatories would require approximately
  (10 x 150 kW + 1 x 500 kW) x 24 h/day = 4.8e4 kWh/day,
which, divided among 7000 US astronomers, is 7 kWh/day/astronomer.
```

Large radio arrays require substantially more energy than large optical telescopes. Approximately 50% of this energy is used to drive the antennas and operate the receivers, while the rest is used for other array operations (hardware correlators, control computers, personnel facilities, etc.).[11] Again, we chose two such toy model arrays to account for the larger number of smaller facilities in operation.

```
The Very Large Array reported a total energy budget of slightly more than 9e6 kWh in 1998.[11]
This implies a daily consumption of (9e6 kWh/yr / 365 days/yr) = 2.5e4 kWh/day,
or approximately 4 kWh/day per astronomer.
Operating two such arrays therefore contributes 8 kWh/day/astronomer.
```

**Supercomputing Centers:** There are a handful of large common-user supercomputing facilities in operation in the US. How much of the nation's supercomputing power is harnessed to perform astrophysical computations?

```
Under the Department of Energy's INCITE program,[12] astronomy- and astrophysics-related projects
received 5e7 processor-hours in 2008.  Taking into account the different power consumption levels
of the various supercomputing facilities,[13] these projects consumed a total of 1.5e6 kWh, or
~1 kWh/day/astronomer, averaged over the entire community for the year.
```

**2.4 Travel**

Astronomers travel a lot. Some (the observers, in taking data) travel more than others (the theorists), but all attend highly valuable workshops and conferences, and forge collaborations and spread understanding by giving seminars at other institutions. The distribution of annual mileages will have quite a long tail, with a small number of senior astronomers covering very large distances as they sit on various political, organizational, and telescope time-allocating committees. We have very little data on the distribution of annual astronomer flight distances, but adopt the following realistic scenario for our toy model of energy consumption. Suppose that an average astronomer's yearly itinerary consisted of 1 round-trip flight each from Los Angeles[14] to Washington DC (for the AAS meeting, 4600 miles), Hawaii (for an observing run, 5100 miles), Seattle (to give a seminar, 1900 miles), and Paris (to an international conference, 11300 miles), for a total of around 23000 air miles. (Note that the mean distance traveled *for work purposes* by the contributing authors of this paper is consistent with this estimate.)

```
Using the airplane energy consumption rate from the introduction above,
we estimate that the average astronomer uses, through flying,
  23000 miles/yr x 1.8kWh/mile / 365 days/yr ~ 113 kWh/day/astronomer
```

In terms of miles flown, astronomers are comparable to typical business travelers: according to the National Business Travelers' Association, the average traveling business person flies approximately 24000 miles per year.[15][16]

**2.5 Discussion**

As the reader may have noticed, it is not so easy to compile accurate estimates of energy consumption. It is much easier to find out the financial cost of a project (in USD) than its true cost (in kWh)! Nevertheless, even from our highly-simplified picture, we can see certain economies of scale in play: operating large ground-based facilities is not as energy-consuming as one might have expected, when averaged over the whole astronomical community. Despite its simplicity, our toy model does allow us to draw quite a strong conclusion: if we are looking to reduce our energy consumption, we can have by far the biggest impact by **decreasing the amount of energy we spend on flights.**

```
Adding up the approximate contributions we have identified above in our toy model,
we estimate the additional total energy consumption of the average US astronomer at work to be
  (4+7+8+1+113) ~ 133 kWh/day/astronomer.
```

Astronomy is a small profession. What impact are we having on US national energy consumption?

```
Taking the population of the US to be 300 million, we find that astronomy costs
   133 kWh/day/astronomer * (7000 / 3e8) = 0.003 kWh/day per US citizen.
```

The cost of doing astronomy is a miniscule proportion (0.0012%) of the total US energy budget. **We conclude that the energy price per citizen of doing astronomy is negligibly small.** The entire daily energy consumption of US astronomy is roughly equivalent to that used by the 1.6 million motorists of Los Angeles in half a minute of their daily commutes.[17]

## 3. Recommendations

The above analysis represents the first order-of-magnitude estimate of the contribution of the U.S. astronomical community to national energy consumption. **We recommend a more thorough investigation by a panel of members from different astronomical disciplines to ensure that a higher level of accuracy is achieved on these estimates before any implementation plan is formulated.**

The purpose of this paper is to motivate the astronomical community to analyze how we consume energy and to find ways to reduce this consumption. It is not the product of astronomy using a particularly large amount of energy, but is instead compelled by the realization that all areas of society (public, private, personal, and professional) should seek to understand and reduce their energy consumption. In the remainder of the paper, we make and discuss specific recommendations for actions that should be taken either by individual astronomers or their funding bodies in order for us to make progress towards this goal. At each stage, we point out ways in which the behavior of astronomers in their professional capacity might have considerably wider impact, if properly publicized, simply as a result of the public's great interest in our field.

**3.1 Improving the Efficiency of Travel in Astronomy**

The single biggest effect the average astronomer can have on his or her professional energy consumption is to reduce the number of miles flown in airplanes. This is problematic: experience shows that much of this travel is very important for successful research programs. How can we reduce our energy consumption without damaging our science?

**Observing:** Many large observatories now operate in service mode, rather than hosting guest observers. In fact, service mode and queue mode observing are now standard at some of the largest optical telescopes such as the VLT and Gemini Observatory and, of course, all space-based telescopes, where they have proven to be very successful. While this model was primarily implemented to maximize the science-efficiency,[18][19] it is of course also far more energy-efficient. A new generation of completely robotic telescopes have come online in the last decade: these are also very energy-efficient. We anticipate most future observatories being designed to operate in some kind of remote mode.

Where it is necessary for an astronomer to be on-hand while observing, remote observing systems can be implemented and made to work at high science productivity. In some cases this will involve significant investment in high bandwidth, stable video link infrastructure either at the guest observers' institutions or at some common local facility. For example, this approach has been adopted very successfully during the last decade by the Caltech and University of California astronomy groups for using the Keck Observatory.[20] We recommend that any future systems like this be funded not only enthusiastically but also visibly; the accompanying publicity must advertise the reasoning behind the upgrade, namely that it is important to reduce the distances we all travel, whenever possible, in order to save energy. We note that production and installation of such technology can be performed by private sector contractors, who can be encouraged to use their association with astronomy when selling their products to businesses and industries. Once installed, these remote observing systems can also be used for video-conferencing (see below).

**Conferences and meetings:** Conferences and workshops are extremely productive research activities - and it is often the informal connections made between researchers "offline" that lead to the most successful new collaborations and projects. Whatever steps we take to reduce energy consumption in this arena, we must be careful to preserve these high-value interactions. The output of other meetings - of committees, collaborations and working groups - are often less dependent on face-to-face interaction. These lower-value interactions are the ones where we can afford to compromise more when pursuing the reduction of energy consumption. For example, the Fermi Gamma-ray Space Telescope collaboration holds 6-7 virtual meetings per week, attended by scientists from many countries and time-zones.

How can we reduce our energy consumption while retaining the high scientific value of cross-collaboration conferences? Airplanes and automobiles are comparably energy-hungry; buses and trains are an order of magnitude more efficient.[5] Organizing meetings close to concentrations of astronomers, or in places with good train links to other cities, would reduce the energy costs of these meetings. This solution may become even more workable as the US moves towards mass ground transit in the next decade. Moreover, a visible shift towards the use of buses and trains could have very high publicity value, since the biggest single contribution to the energy consumption of the average US citizen (section 2.1) is traveling by automobile. Astronomers are fortunate in that much of their work (reading, writing, programming) can be done on a laptop: with sufficient planning, train travel does not involve lost research time. We also suggest conference funding models where a flat registration fee is charged, and participants then apply for travel grants: this could allow the conference organizers to enforce energy-efficient travel, and would equalize costs to prevent the few distant participants being discouraged from attending.

It is even possible to avoid traveling to a conference completely, while retaining its high-value scientific interactions. Video-conferencing systems now allow for remote participation at a somewhat reduced level, but have considerable potential. They are expensive and perhaps not obviously covered by many astronomical grants, but we believe that they ought to be. Exploring ways to enhance the quality of remote scientific interactions to the point where they replicate physical attendance as much as possible is clearly a worthwhile activity. One suggestion is to investigate the possibility of holding astronomy conferences in "virtual worlds".[21] This technology is still being developed, but is already mature enough for us to be organizing small workshops[22] or virtual institutions. For example, the Meta Institute for Computational Astrophysics[23] exists almost entirely in a virtual world, with scores of members giving presentations at regularly-organized meetings. Given the high level of interest in new astronomical results, the coverage of press releases held in such virtual worlds would have more impact on the public's energy awareness than other events held there might.

To make these systems work for astronomers may well require some investment in the development of the software, if it is found that astronomers have particular needs that the general packages cannot support. For example, in order for astronomers from the entire global community to be able to participate in these virtual conferences, the software must be made available internationally, and be able to function across all platforms, and across a wide range of available connection bandwidths. This may be a particularly important issue for participation by astronomers in the developing world, whom we must take care not to exclude in any way in the name of energy efficiency.

**We recommend that funding agencies invest in the development and support of telecommunication and software systems that lead to reduced energy consumption through reducing distance traveled.**

**We also recommend that conference organizers and their funding bodies take the energy budget of their meeting into account when planning the meeting, including providing incentives for ancillary workshops and splinter meetings that will decrease the future travel needs of participants; this will maximize valuable scientific interaction while minimizing total energy consumption.**

### 3.2 Making Infrastructure Energy-efficient

Astronomy is one of the avenues through which humanity obtains a better understanding of nature: it is our responsibility to harness nature in the most sustainable way through using energy-efficient facilities. The price of energy is expected to rise in the coming decade: making astronomical infrastructure energy-efficient may well pay for itself despite the significant costs up front.

Many government buildings and university departments might be expected to be overhauled or replaced in the coming decade, making them more energy efficient. Certainly such schemes form part of the American Recovery and Reinvestment Act[24] economic stimulus package. In this section we focus on the more visible astronomical facilities, whose design, when publicized, can be used to promote the need for energy-efficient infrastructure everywhere.

**Observatories:** Telescopes are often built in energy-rich environments: in deserts or on the tops of mountains where, by choice, the skies are clear. Such places make excellent sites for solar power plants; wind power will also often work well there, and can be combined with solar paneling. We can imagine supplying an observatory with all its electricity from such sustainable sources. Telescope operation at night would then need significant power to be drawn from batteries or other storage devices. How big a solar installation would be required to supply the 150kW required by one of our fiducial optical observatories?

```
Using solar generators operating at (on average) 30 W/m^2,
the required area is 1.5e5 / 30 = 5000 sq m, about the size of a football field.
```

Solar power plants with comparable efficiency have been in use in the Mojave Desert for more than 20 years.[25] The construction of such a power plant next to the observatory would not be cheap (although for new observatories a national grid connection may come at a significant cost as well). However, the publicity value of such an enterprise would be very high: financially-expensive solar power will take some getting used to, and awareness needs to increase. We can anticipate some tension between the protection of areas of cultural and ecological significance, and the need to provide, and be seen to provide, clean energy; this is a general problem that we will all need to address in the next decade. There are at least two small observatories already 100% powered by renewable energy: the Observatorio Cerro Armazones in Chile[26] and the Kielder Observatory in the UK.[27] Moreover, other plans are being made: for example, the Square-Kilometer Array South Africa is considering obtaining at least some of its power from solar panels.[28]

**Data and computing centers:** These facilities are not necessarily built close to high energy-density deserts; nevertheless, sustainable sources of energy for them could be found. Google Inc., with its prodigious computing needs, has embarked on an ambitious initiative to power their headquarters with solar power.[29] However, if the most cost-effective way of providing sustainably-generated power to computing centers is to generate it elsewhere (such as in a desert) and then transfer it via the national grid, then it would seem appropriate to invest in such remote power stations as part of the computing center construction process. The publicity value of this is that it makes explicit the link between the activity and the need for a renewable source of energy to power that activity: as a society we have come to take electricity generation for granted.

There are also other ways in which supercomputing centers can be made energy-efficient. For example, the LEED-certified Illinois Petascale Computing Facility (supported by NSF) will use innovative ways to improve the efficiency of the processor cooling, and avoid the need for inefficient backup power generation.[30] One can also imagine using the heat from the computer racks in supercomputer centers to heat water for the building, and for surrounding homes.

Future astronomical facilities should certainly be developed with a goal of achieving LEED certification. Moreover, facilities like observatories and computing centers are extremely impressive feats of engineering, whose functions are very effective at capturing the public's attention and imagination. This should be capitalized upon.

**We recommend that the astronomy funding agencies seize the opportunity to make astronomical facilities energy-efficient, and then promote them as examples for the rest of society.**

**3.3 Energizing Astronomers, and Publicizing their Activities**

The benefits of implementing any of our recommendations can be amplified greatly by publicizing them effectively. If true-cost accounting, implemented at the highest level, is not adopted sufficiently soon, then we can still hope that US residents will act to avert the energy crisis they are facing by changing their lifestyles and working practices to more energy-efficient ones. This is of course very difficult to ask for in such hard economic times; however, there may be some changes that Americans and their businesses can take that save both money and energy, at the cost of some convenience. One example is the use of mass transit when commuting to work. Another is replacing flying to business meetings with videoconferences. Another may be making their buildings more energy-efficient. All of these are changes we have recommended for astronomers, and all are changes that may be met with resistance. We argue that a social revolution is needed, towards a more energy-conscious and then energy-efficient way of living and working, and that astronomers can play a leading role in this by virtue of their visibility in the media.

The IAU's goals for the International Year of Astronomy 2009 include "bring[ing] the issues of natural environment and energy preservation to the agenda of decision makers."[31] We hope this spirit continues into the next decade, with astronomers playing a role in bringing the issue of energy efficiency to the attention of the public. Doing this will require some resources: we can imagine agencies like the NSF and the AAS assigning manpower to specifically facilitate this publicity. A good example of the way in which astronomy can help raise the profile of energy-efficient technology is the "BLOOMhouse" installation at McDonald Observatory.[32][33] Visitors to the observatory can see for themselves just how energy-efficient buildings can be, after learning about the power of the sun in an astrophysical context. The new Caltech Cahill Center for Astronomy & Astrophysics is another example of a LEED-gold certified facility which merges sustainable building practices and materials with an energy efficient design.

We would also encourage more informal attempts by astronomers to publicize their energy consumption, and their quantified understanding of it, to their local public. To see how this would work, we can consider the "reach" of astronomers as follows. There are a number of online communities actively interested in astronomy. How big are these? The Google Earth community has about a million members: the forum post "Where is Hubble Now?"[34] which was written by a professional astronomer, was viewed more than 64,000 times. The number of citizen scientists who have classified galaxy morphologies on the Galaxy Zoo[35] website is now greater than 150,000. Many people can also be reached through astronomical textbooks. How about more direct means of contact?

```
Suppose that 500 of the 7000 US astronomers (~10%) are faculty members actively participating in
outreach work. In a given year, suppose they lecture to some 400 students (the majority of whom
are not majoring in astronomy), and give a public talk to an audience of 50.
The total combined audience for all these lecturers is 500 x (400 + 50) = 225000
```

Just by giving talks, the astronomy community can influence a group of people roughly 30 times larger than itself. A social epidemic needs an efficient network: we sincerely hope that astronomers will be key nodes in this network. As the UN HDR report[3] concludes, "The world lacks neither the financial resources nor the technological capabilities to act. What is missing is a sense of *urgency, human solidarity and collective interest.*"

**We therefore recommend that astronomers combine the presentation of their exciting science with explanations of and motivations for their efforts to reduce their energy consumption while carrying out that science.**

---

**References:**

1. Climate Change 2007: The Fourth Assessment Report of the IPCC (http://www.ipcc.ch/ipccreports/ar4-wg2.htm)
2. Sokolov, et al. (2009) (http://globalchange.mit.edu/resources/gamble/)
3. UN Human Development Report (2007/2008) (http://tinyurl.com/5nthf8)
4. American Astronomical Society Endorsement of AGU Statement on Climate Change (http://tinyurl.com/bxwuzf)
5. MacKay (2008): Sustainable Energy without the Hot Air (http://withouthotair.com)
6. UN Human Development Report (2007/2008) (http://hdrstats.undp.org/countries/data_sheets/cty_ds_USA.html)

7. American Astronomical Society Membership Count (http://aas.org/membership/memcount.php)

8. EPA: Greenhouse Gas Emissions from a Typical Passenger Vehicle (http://tinyurl.com/57qfcq)

9. Bureau of Transportation Statistics (http://tinyurl.com/6j7e9d)

10. SOFIA "Aircraft Facts" (http://www.sofia.usra.edu/Sofia/aircraft/aircraft_facts.pdf, see also http://tinyurl.com/c5qrbp)

11. VLA Test Memo # 223 (http://www.vla.nrao.edu/memos/test/223/223.pdf)

12. U.S. Department of Energy, INCITE Awards (2008) (http://www.sc.doe.gov/ascr/INCITE)

13. The Green500 List: Ranking of the most energy-efficient supercomputers in the world (http://www.green500.org)

14. Los Angeles Almanac (http://www.laalmanac.com/transport/tr53.htm)

15. GreenBiz: Greening Your Company's Business Travel (2008) (http://tinyurl.com/d4c9x4)

16. The 2008 NBTA CSR Toolkit: A Primer on Responsible Travel Management (http://tinyurl.com/bvxrs9)

17. Los Angeles Dept. of Transportation, "2008 Transportation Profile" http://ladot.lacity.org/pdf/PDF10.pdf

18. Gemini Focus: Observing Efficiency at Gemini Observatory (December 2005; page 54) (http://tinyurl.com/cjfdmp)

19. VLT Service Mode Philosophy (http://www.eso.org/sci/observing/phase2/SMPhilosophy.html)

20. W. M. Keck Observatory: Policy on Mainland Observing (http://tinyurl.com/abqgpz)

21. e.g. Second Life (http://secondlife.com)

22. Riley (2008): If A Conference Is Held In Second Life, Will Anyone Listen? (http://tinyurl.com/yscs2n)

23. Meta Institute for Computational Astrophysics (http://tinyurl.com/arpexu)

24. American Recovery and Reinvestment Act (2009) (http://www.recovery.gov)

25. Kearney (1989): IEEE Power Review (http://ieeexplore.ieee.org/iel1/39/2848/00087383.pdf?arnumber=87383)

26. Observatorio Cerro Armazones: A 100% "Green" Observatory (http://www.astro.ruhr-uni-bochum.de/astro/oca/green.html)

27. Kielder Observatory (http://www.kielderobservatory.org)

28. Square-Kilometer Array South Africa, Newsletter (May 2008) (http://www.ska.ac.za/newsletter/issues/08/02.html)

29. Google Solar Panel Project (http://www.google.com/corporate/solarpanels/home)

30. Illinois Petascale Computing Facility: Blue Waters project (http://www.ncsa.uiuc.edu/BlueWaters/pcf.html)

31. International Year of Astronomy 2009: Goals and Objectives (http://www.astronomy2009.org/general/about/goals/)

32. BLOOMhouse, University of Texas, (http://soa.utexas.edu/solard)

33. McDonald Observatory News Releases (April 2008) (http://mcdonaldobservatory.org/news/releases/2008/0422.html)

34. Conti (2007), Google Earth community: Where is Hubble now? (http://tinyurl.com/d3p7xq)

35. Galaxy Zoo (https://www.galaxyzoo.org)

---

**Acknowledgments:**

Many thanks to Jennifer Mehl of UCSB physics department computing services for getting the wiki-page (http://low-energy-astro.physics.ucsb.edu) used to write this paper set up on a very short timescale. We are also grateful for all the feedback we received from the community when writing this paper, especially from those who agree with its sentiments but who could not, for whatever reason, act as its co-authors.